\documentclass[aps,prd,amsmath,amssymb,superscriptaddress,nofootinbib,10pt,showpacs,twocolumn]{revtex4-2}

\usepackage[]{latexsym}
\usepackage[english]{babel}
\usepackage{graphicx}
\usepackage{hyperref}
\usepackage{color}
\usepackage{bm}

\usepackage{array} 

\usepackage{orcidlink}

\allowdisplaybreaks

\newcommand{\sumnu}{M_\nu}

\begin{document}

\title{Unveiling Neutrino Masses: Insights from Robust (e)BOSS Data Analysis and Prospects for DESI and Beyond}
\author{{Hernán E. Noriega}\orcidlink{0000-0002-3397-3998},}
\email{henoriega@icf.unam.mx}

\affiliation{Instituto de Ciencias F\'isicas, Universidad Nacional Autónoma de México,  62210, Cuernavaca, Morelos.}
\affiliation{Instituto de Física, Universidad Nacional Autónoma de México, Apdo. Postal 20-364, 01000, D.F, México.}

\author{{Alejandro Aviles}\orcidlink{0000-0001-5998-3986},}
\email{aviles@icf.unam.mx}

\affiliation{Instituto de Ciencias F\'isicas, Universidad Nacional Autónoma de México,  62210, Cuernavaca, Morelos.}


\begin{abstract}
Recent findings from DESI BAO combined with Planck CMB data have set an upper limit on the total neutrino mass of $\sum m_\nu < 0.072\, \text{eV}$ (95\% confidence level), ruling out the inverted hierarchy. Indeed, methods that rely on the background expansion of the Universe tend to suggest negative neutrino masses. In this work, we contribute to the quest for accurately constraining neutrino mass using cosmological probes. By conducting a full-shape analysis on data from BOSS, eBOSS, and synthetic power spectra, we showed that projection effects can significantly influence constraints on neutrino mass, rendering these measurements largely unreliable. Our results highlight the need for better techniques to measure the neutrino mass accurately. Based on the large-scale structure suppression, we identified a critical blind spot in the full-shape analysis. By splitting the galaxy power spectrum into broadband and wiggles, we noticed that information on neutrino mass is primarily extracted from the suppressed wiggles rather than broadband suppression. This opens the possibility of developing alternative methods based only on the wiggles of the power spectrum that can be more robust than those heavily reliant on background evolution.
\end{abstract}

\maketitle

\begin{section}{Introduction}
Neutrino oscillation experiments have accurately measured the squared-mass differences between their definite mass states \cite{Esteban:2020cvm}, implying that there are at least two states of massive neutrinos. These states can be ordered either in a normal hierarchy (NH), where the sum of their masses satisfies $M_{\nu} \equiv \sum_i m_{\nu,i} > 0.058 \, \text{eV}$, or in an inverted hierarchy (IH), with $M_{\nu} > 0.10 \, \text{eV}$. On the other hand, the most recent results from the Karlsruhe Tritium Neutrino (KATRIN) experiment have put an upper bound of $0.45\, \text{eV}$ at the $90\,\%$ confidence level (c.l.) on the lightest state \cite{Aker:2024drp}.

However, the most constraining upper bounds on the neutrino masses so far have come from cosmology \cite{Planck:2018vyg,DESI:2024mwx}, although indirectly, by measuring their gravitational influence over the expansion history of the Universe and the large-scale structure (LSS); see \cite{Lesgourgues:2006nd, Wong:2011ip} for reviews on neutrino physics in cosmology.
Recently, the Dark Energy Spectroscopic Instrument (DESI) put a very tight constraint on the sum of the neutrino masses, $M_\nu<0.072 \,\text{eV}\,\,(95\,\%\,\text{c.l.})$, by combining data from the cosmic microwave background radiation (CMB) and DESI BAO \cite{DESI:2024mwx}, assuming the flat $\Lambda$CDM model. However, when considering evolving dark energy with the CPL (or $w_0$--$w_a$) parametrization, the mass constraint is severely relaxed to $M_\nu<0.195 \,\text{eV}$ \cite{DESI:2024mwx}. Despite the success of such analyses, the neutrino effect is very mild, and even a small discrepancy in the measurements of the other cosmological parameters can yield misleading results due to degeneracies.

An aspect that has attracted attention is that the posterior profile obtained from CMB+DESI BAO data peaks at $\sumnu=0$, exhibiting a behavior akin to the tail of a distribution that finds its best-fit at negative neutrino masses. 
This is not new, but was observed using CMB data alone \cite{Planck:2018vyg}, which further constrains $\sumnu < 0.24 \,\text{eV}\,\,(95\,\%\,\text{c.l.})$, that is improved to $\sumnu < 0.12 \,\text{eV}\,\,(95\,\%\,\text{c.l.})$ by adding BAO data from the Baryon Oscillation Spectroscopic Survey (BOSS) DR12 galaxies \cite{BOSS:2016wmc}. The peak profile behavior has also been observed using measurements of clustering involving galaxies, quasars, and the Lyman-$\alpha$ forest from the completed Sloan Digital Sky Survey (SDSS) together with CMB data \cite{eBOSS:2020yzd}. 
The reason behind this behavior in the posterior profile is unclear. A recent study argues that the peak at $\sumnu=0$ and the remarkably tight constraints imposed by DESI, nearly crossing the threshold $\sumnu<0.06 \,\text{eV}$ at $2\, \sigma$, could be a signal of new physics in the neutrino sector \cite{Craig:2024tky}. Similarly, this behavior could also point to modified gravity, which often exhibits degeneracy with massive neutrinos \cite{Aviles:2021que,Moretti:2023drg}.

In this work, we utilize full-shape modeling of the galaxy power spectrum to revisit the constraints on the total neutrino mass using LSS data from BOSS DR12 galaxies \cite{2013AJ....145...10DBOSS:2016wmc, BOSS:2016wmc} and extended BOSS (eBOSS) quasars \cite{Dawson:2015wdb, eBOSS:2020yzd}. Our goal is to understand the sources that could lead neutrino mass constraints to favor negative values. Specifically, this work addresses the following questions: Is the peak at $\sumnu  = 0$ also seen in full-shape analyses of (e)BOSS data? Are the constraints on neutrino mass affected by projection effects? If so, can the inclusion of more data help alleviate these effects?
To this end, we analyze individual redshift bins of BOSS and eBOSS, as well as the combined datasets. The analysis shows that the well-marked peak at $\sumnu  = 0$ is driven by eBOSS, which surprisingly places tighter constraints on the neutrino mass than the two non-overlapping redshift bins of BOSS, $z_1$ and $z_3$, together. This behavior persists even with the inclusion of more information through external priors. To gain more insight, we construct noiseless synthetic data generated with the flat $\Lambda$CDM model using \textit{Planck 2018} best-fit values for cosmological parameters, including $\sumnu = 0.06 \,\text{eV}$. In both cases (using real or synthetic data), we obtain very similar profiles for the neutrino mass posteriors. This suggests that projection effects (also known as prior volume effects) significantly impact neutrino mass constraints, we verify their presence by comparing frequentist and Bayesian approaches.

Given the reasons stated above, it is worthwhile to explore alternative approaches for robustly extracting information on neutrino mass. In particular, we are interested in approaches that primarily rely on the suppression of the galaxy power spectrum, and not on the background evolution. By splitting the power spectrum into BAO wiggles and broadband contributions, we show that the information on the sum of the neutrino masses is primarily extracted from the former, contrary to common belief. This discovery opens the possibility of constructing methods that rely solely on extracting the neutrino mass from the relative amplitude of the BAO wiggles. Although one expects that such methodologies would provide looser constraints, they could be more robust against changes in the background expansion of the Universe.

The rest of the paper is organized as follows. In sect.~\ref{sect:theory} we present the full-shape modeling we use in this paper; in sect.~\ref{sect:data} we show the datasets, likelihood and priors used throughout; in sect.~\ref{sect:analysis} we present our numerical results; in sect.~\ref{sect:wiggles}
we delve into the wiggle-broadband decomposition. Finally, we present our conclusions and final remarks in sect.~\ref{sect:conclusions}.
\end{section}

\begin{section}{Theory}\label{sect:theory}
The Perturbation Theory (PT) and Effective Field Theory (EFT) for LSS has been developed over the past decades by several authors, e.g. \cite{Bernardeau:2001qr,McDonald:2009dh,Baumann:2010tm,Vlah:2015sea}, and has recently proven useful for directly fitting the galaxy power spectra \cite{Ivanov:2019pdj,DAmico:2019fhj} (see also 
\cite{Wadekar:2020hax,Chudaykin:2020aoj,Philcox:2021kcw,Philcox:2022frc,Tanseri:2022zfe,Nishimichi:2020tvu,Chen:2020zjt,Tsedrik:2022cri,Carrilho:2022mon,Nunes:2022bhn,Ramirez:2023ads}). Since then, a few codes has been released to compute the galaxy power spectrum in redshift space: including \texttt{class-pt} \cite{Chudaykin:2020aoj}, \texttt{velocileptors} \cite{Chen:2020fxs,Chen:2020zjt}, \texttt{PyBird} \cite{DAmico:2020kxu}, \texttt{CLASS-OneLoop} \cite{Linde:2024uzr}, all of them utilizing Einstein-de Sitter (EdS) kernels. In this work we compute the galaxy power spectrum using the code \texttt{FOLPS}\footnote{Fast 1-loop power spectrum (\texttt{FOLPS}). Publicly available at \url{https://github.com/henoriega/FOLPS-nu}. JAX implementation: \url{https://github.com/cosmodesi/folpsax}.} \cite{Noriega:2022nhf}, which is based on the PT/EFT in the presence of massive neutrinos developed in \cite{Aviles:2020cax,Aviles:2021que,Noriega:2022nhf}, and employs the perturbative \texttt{fk}-kernels, as named in \cite{Aviles:2021que,Rodriguez-Meza:2023rga}. These kernels maintain the scale-dependence of the growth factors $f(k,t) = d \log D_+(k,t) / d \log a(t)$, which arise from the introduction of the neutrino mass as a new scale into the theory.  
The use of these kernels has been shown to provide more stringent constraints on the sum of neutrino masses than the EdS ones under certain settings \cite{Noriega:2024eyu}. Moreover, the computation time of loop corrections is very fast, ensuring both efficiency and improved accuracy \cite{Noriega:2022nhf}, making \texttt{fk}-kernels a highly attractive alternative.

We use the galaxy power spectrum \cite{Kaiser:1984sw,Scoccimarro:2004tg,Taruya:2010mx,Aviles:2020wme,Aviles:2021que,Perko:2016puo,Chen:2020zjt,Vlah:2018ygt}
\begin{align}\label{PS_EFT}
P^\text{EFT}_s(k,\mu) &= P_{\delta\delta}(k) + 2 f_0 \mu^2 P_{\delta\theta}(k) + f_0^2 \mu^4 P_{\theta\theta}(k) \nonumber\\
&\quad + A(k,\mu) + D(k,\mu) \nonumber\\
&\quad + (\alpha_0 + \alpha_2 \mu^2 + \alpha_4 \mu^4 +\alpha_6 \mu^6) k^2 P_L(k) \nonumber\\ %
&\quad + \alpha^{\rm shot}_0 + \alpha^{\rm shot}_{2} (k\mu)^2.
\end{align}
The first line in this equation is the non-linear Kaiser power spectrum \cite{Kaiser:1984sw,Scoccimarro:2004tg}. 
The function $A(k,\mu)$ was introduced in \cite{Taruya:2010mx} and is obtained from correlators of three fields, either density or velocity fields, while function $D(k,\mu)$ correspond to correlators of four fields. Expressions for these functions can be found in \cite{Aviles:2020wme,Noriega:2022nhf}. The last two lines include the EFT countertems $\alpha_{0,2,4,6}$ and stochastic noise parameters $\alpha^{\rm shot}_{0,2}$.  In this work we use the third order bias expansion of \cite{McDonald:2009dh}, which consider bias parameters $b_1$, $b_2$, $b_{s^2}$ and $b_\text{3nl}$.\footnote{Alternatively, \texttt{FOLPS} can use the bias expansion with $b_{\mathcal{G}_2}$ and $b_{\Gamma_3}$ \cite{Assassi:2014fva} instead of $b_{s^2}$ and $b_\text{3nl}$.} We refer the reader to \cite{Noriega:2022nhf,Rodriguez-Meza:2023rga} for details in all the expressions of the theory for the beyond EdS kernels.

Before moving to fitting the neutrino mass, we want to compare the performance of \texttt{FOLPS} with previous BOSS analyses, within the framework of the flat $\Lambda$CDM model, assuming a fixed neutrino mass, $M_\nu = 0.06\, \text{eV}$. In Fig.~\ref{fig:comparison_LCDM} we show such comparisons to \texttt{PyBird}  \cite{DAmico:2019fhj} (D'Amico et al.~2020), \texttt{class-pt} \cite{Ivanov:2019pdj} (Ivanov et al.~2020) and \texttt{velocileptors} \cite{Chen:2021wdi}  (Chen et al.~2022). Discrepancies in the results mainly stem from variations in nuisance parameters and their priors.\footnote{In \cite{Simon:2022lde} it is shown that most of the differences between \texttt{PyBird} and \texttt{class-pt} results previously observed in \cite{Nishimichi:2020tvu} arise from weight effects due to the choice of EFT counterterms priors.}

Recently, the debate on EFT model comparisons has concluded, demonstrating that different EFT codes yield nearly identical constraints under equivalent settings (nuisance parameters, priors, etc).  This was shown in  \cite{Maus:2024sbb} using \texttt{FOLPS}, \texttt{velocileptors}, and \texttt{PyBird} on LRG, ELG, and QSO mocks, as well as noiseless synthetic data generated by EFT models, even for mocks with extremely large volumes.

Finally, it is interesting to note from Fig.~\ref{fig:comparison_LCDM} that all BOSS analyses presented with colored dots and error bars reach lower values on $A_s$ compared to Planck. This behavior is well-known, see e.g. \cite{Troster:2019ean, DiValentino:2020vvd, Abdalla:2022yfr, Chen:2024vuf}. However, constraints on $A_s$ can be improved by sampling bias parameters scaled by the amplitude of matter fluctuations $\sigma_8$, rather than directly sampling bias parameters \cite{Maus:2024dzi}. The results of using scaled biases with $\sigma_8$, indicated by red stars and error bars, are consistent with Planck  within $1\, \sigma$. We will revisit this point in Sect.\ref{sect:analysis}.

\begin{figure}
    \includegraphics[width=3.3 in]{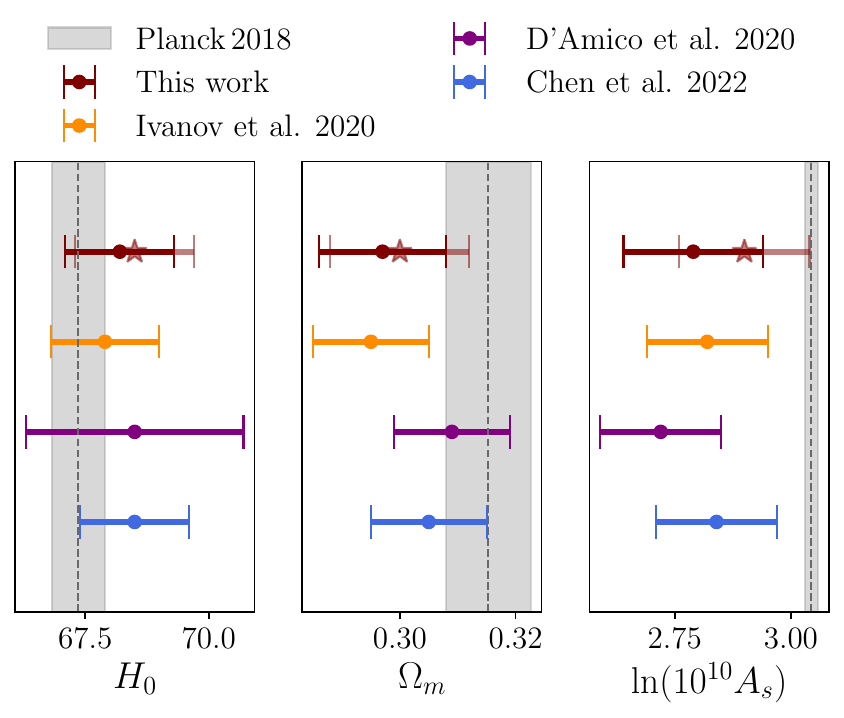}
 	\caption{ Comparison of 1-dimensional credible intervals (means and 68\,\% c.l.) from full-shape analyses of the BOSS power spectrum using different pipelines \cite{Ivanov:2019pdj, DAmico:2019fhj, Chen:2021wdi}, assuming a flat $\Lambda$CDM model with $M_{\nu} = 0.06\, \text{eV}$.
    Points (stars) with error bars represent results from sampling bias parameters directly (scaled with powers of $\sigma_8$).
    The gray regions (dashed lines) correspond to the 68\% c.l. (best-fit) of Planck 2018 \cite{Planck:2018vyg}. 
    } 
   \label{fig:comparison_LCDM}
 \end{figure}
\end{section}

\begin{section}{Data and likelihood}\label{sect:data}
In this section we briefly introduce the datasets \cite{Beutler:2021eqq} and the analysis setup \cite{Noriega:2024eyu} used throughout this work. 

\begin{subsection}{Datasets}
We analyze the clustering of galaxies drawn from the publicity available BOSS DR12 galaxy sample \cite{2013AJ....145...10DBOSS:2016wmc, BOSS:2016wmc} and eBOSS DR16 quasar sample \cite{Dawson:2015wdb, eBOSS:2020yzd}, which are part of the SDSS-III and -IV projects, respectively.
For the BOSS galaxy sample, we concentrate on the two non-overlapping redshift bins $0.2 < z_1 < 0.5$ and $0.5 < z_3 < 0.75$ with effective redshifts $z_{\rm eff} = 0.38$ and $z_{\rm eff} = 0.61$, respectively. At high redshifts, we limited our analysis to the eBOSS quasars dataset, covering the range $0.8 < z < 2.2$ with an effective redshift of $z_{\rm eff} = 1.52$.

We use the products (data power spectra multipoles, window functions, and covariance matrices) provided in \cite{Beutler:2021eqq}.\footnote{\url{https://fbeutler.github.io/hub/deconv_paper.html}} The measured power spectra multipoles were computed assuming a \textit{fiducial} flat $\Lambda$CDM cosmology with $\Omega^{\rm fid}_m = 0.31$, introducing artificial anisotropy in the galaxy clustering. This anisotropy is accounted for through the Alcock-Paczynski (AP) effect (see e.g., Section 5.1 from \cite{Noriega:2024eyu}). Besides the AP effect, the survey geometry also impacts the measured power spectra. We account for this by convolving the clustering signal with the corresponding window function \cite{Beutler:2021eqq}.
We estimate the numerical covariance matrices from a suite of galaxy mocks, which match the clustering properties and geometry of BOSS and eBOSS samples. Covariances for the BOSS DR12 galaxy sample are extracted from $2\times2048$ realizations of the MultiDark-Patchy mock catalogs for NGC and SGC \cite{Kitaura:2015uqa}. 
Meanwhile, eBOSS DR16 quasar covariances are computed from $2\times1000$ realizations of the EZmocks \cite{eBOSS:2020wwo}, also for NGC and SGC patches.

In addition to analyzing real data from BOSS galaxies and eBOSS quasars, we examine noiseless synthetic data generated from our theoretical model at the same redshifts, using the \textit{Planck 2018} cosmology with $\sumnu = 0.06\, \text{eV}$ as input. The analysis of synthetic data maintains consistency with the real data analysis by employing the same covariance, window function, and settings. 
\end{subsection}

\begin{subsection}{Settings}
Throughout this work, we fit the monopole and quadrupole adopting the wave-number range $0.01 \leq k\, [h\,{\rm Mpc^{-1}}] \leq 0.20$. Our model has been shown to provide unbiased results within this range for mocks with BOSS-like volume \cite{Noriega:2022nhf} and even for extremely large volumes \cite{Noriega:2024eyu, Maus:2024sbb}. Our base setup allows the following parameters to vary:
\begin{align}
&{\rm Cosmological:}\, \{\sumnu,\, h,\, \omega_{\rm b},\,  \omega_{\rm c},\, \ln(10^{10}A_s)\}, \\
&{\rm Nuisance:}\, \{b_1,\, b_2,\, \alpha_{0},\, \alpha_{2},\, \alpha^{\rm shot}_{0},\, \alpha^{\rm shot}_{2}\}.  
\end{align}
The cosmological parameters are varied over sufficiently wide ranges to be considered uninformative, except for the baryons, for which we impose the BBN prior $\omega_b = 0.02237 \pm 0.00037$ \cite{Aver:2015iza, Cooke:2017cwo}. We also keep the spectral index fixed $n_s = 0.9649$ at the best-fit value from \textit{Planck 2018}.
Additionally, we fix the tidal bias and non-local bias according to coevolution theory \cite{Chan:2012jj,Baldauf:2012hs,Saito:2014qha}, using $b_{s^2}= -\frac{4}{7}(b_1-1)$ and $b_{3nl}=\frac{32}{315}(b_1-1)$.
The use of coevolution slightly narrows the posterior distributions, but does not alter our conclusions \cite{Noriega:2024eyu}.

\begin{figure*}
 	\begin{center}
 	\includegraphics[width=6 in]{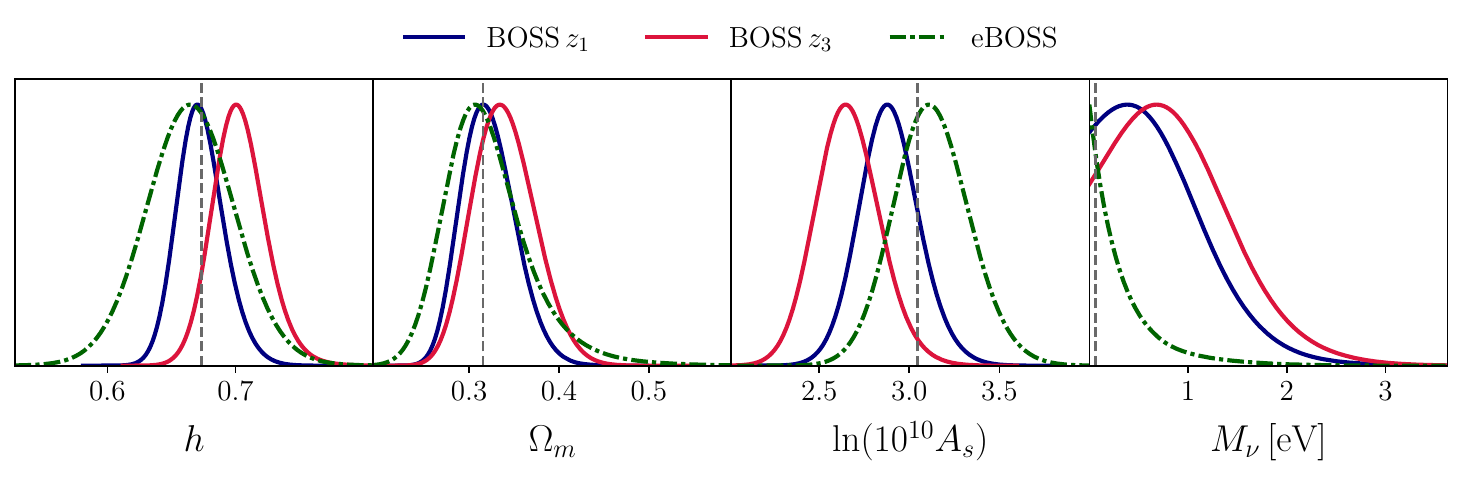}
 	\includegraphics[width=6 in]{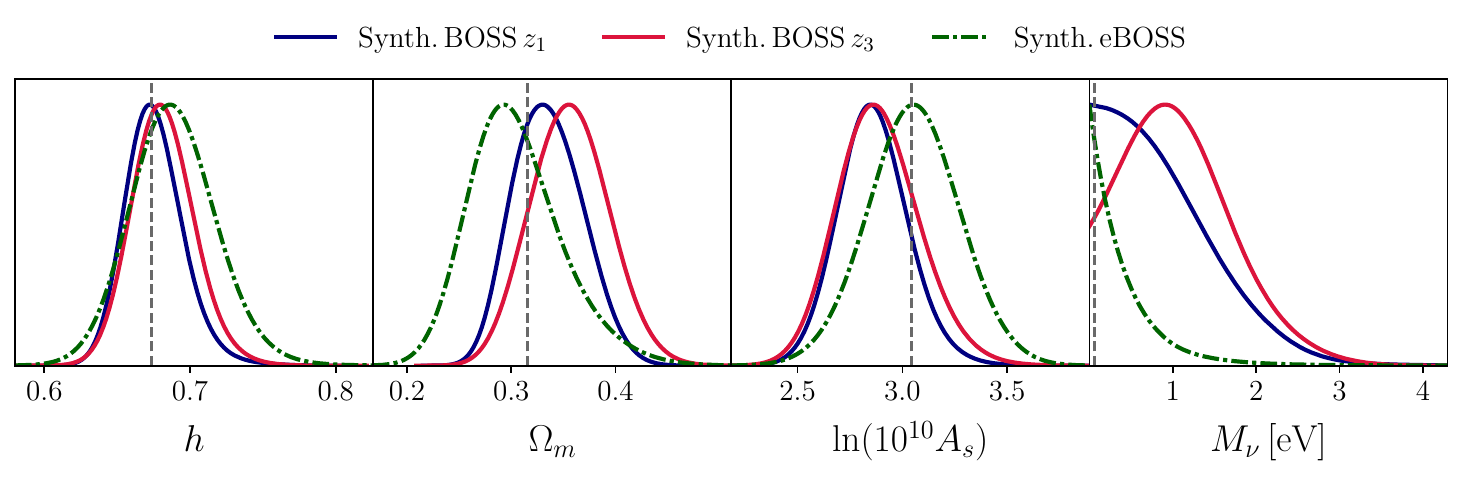}
 	\caption{Constraints from BOSS $z_1$ ($z_{\text{eff}}=0.31$) and $z_3$ ($z_{\text{eff}}=0.61$) galaxies, and eBOSS quasars at redshift $z_{\text{eff}}=1.52$. The \textit{upper panel} shows the results from fitting the real data, while the \textit{bottom panel} shows the results from fitting synthetic noiseless (synth.) data. Vertical dashed lines represent the best-fit values from \textit{Planck 2018}.
  \label{figure:triangular}
  }
 \end{center}
 \end{figure*}

In the Bayesian framework, we analytically marginalize over the EFT and stochastic parameters, as they enter at linear order in Eq.~(\ref{PS_EFT}). The marginalization process sets priors on these parameters; we choose the agnostic option of employing uninformative flat priors on the EFT and stochastic parameters, thereby avoiding potential prior weighting effects from an incorrect choice of these priors \cite{Noriega:2024eyu, Simon:2022lde}. The marginalized Likelihood $\mathcal{L}$ is shown in Eq.~(A.9) of \cite{Noriega:2024eyu}. The complete (unmarginalized) likelihood is employed in the frequentist profile likelihood approach. That is, $\ln \mathcal{L} = - \tfrac{1}{2} \chi^2$, with $\chi^2 \equiv \Delta D^T \text{Cov}^{-1} \Delta D$ and $\Delta D$ representing the residual between the model and the data vectors. In this method, we handle the nuisance parameters directly, as opposed to marginalization, which is commonly used in the Bayesian approach.
\end{subsection}
\end{section}

\begin{section}{Data analysis}\label{sect:analysis}
In this section we present our results, obtained using a flat $\Lambda$CDM model with the total neutrino mass $M_\nu$ as an additional free parameter. Our base model adopts one massive and two massless neutrino states. 
Additionally, we incorporate a BBN prior on the baryon density, otherwise stated, and set the spectral index to Planck's best-fit value, as discussed in Sec.~\ref{sect:data}. The linear $cb$ (cold dark matter + baryon) power spectrum is obtained from the code \texttt{CLASS}\footnote{\url{https://lesgourg.github.io/class_public/class.html}} \cite{Blas:2011rf}, serving as input for \texttt{FOLPS}, which calculates the 1-loop corrections and, subsequently, the galaxy power spectrum multipoles that we compare against the data. To explore the parameter space, we conduct Markov Chain Monte Carlo (MCMC) samplings using the \texttt{emcee} code\footnote{\url{https://emcee.readthedocs.io/}} \cite{ForemanMackey:2012ig}. Contour plots, 1-dimensional posterior plots, and confidence intervals are computed using the \texttt{GetDist} Python package \cite{Lewis:2019xzd}.

Firstly, we independently fit the BOSS $z_1$ and $z_3$ galaxies, as well as eBOSS quasars. The corresponding constraints are presented in the upper panel of Fig.~\ref{figure:triangular}. As expected BOSS, $z_1$ and $z_3$ alone have small constraining power on the neutrino mass,
\begin{align}
&\sumnu< 1.67 \,\text{eV} \quad(95 \,\%, \text{BOSS $z_1$} +  {\rm BBN}),    \\
&\sumnu< 1.94 \,\text{eV} \quad(95 \,\%, \text{BOSS $z_3$} +  {\rm BBN}).     
\end{align}

However, as noticed in \cite{Simon:2022csv}, the upper bounds imposed by eBOSS are unexpectedly stringent,
\begin{align}
&\sumnu< 1.13 \,\text{eV} \quad(95 \,\%, \text{eBOSS} +  {\rm BBN}),    
\end{align}
despite this dataset being the noisiest and having the least amount of data among those utilized in this work. As expected, the rest of cosmological parameters show more relaxed constraints for this dataset. Moreover, the posterior distribution of neutrino mass peaks at $\sumnu=0$, resembling the tail of a truncated distribution that would otherwise peak at negative mass values. This peculiar pattern could be attributed to various factors, ranging from statistical flukes to potential indications of new physics. To ascertain whether this phenomenon persists under controlled conditions, we conduct fits to noiseless synthetic datasets generated with our theoretical model employing the \textit{Planck 2018} best-fit parameters, with $\sumnu = 0.06\, \text{eV}$, while keeping the same covariance matrices and window functions as those used with the real data. Surprisingly, the results demonstrate striking similarities for all cosmological parameters, with the posterior for the neutrino mass exhibiting an almost identical profile and  
\begin{align}
&\sumnu< 1.24\,\text{eV} \quad(95 \,\%, \text{synthetic eBOSS + BBN}).  
\end{align}

This behavior may indicate that we are observing projection effects that drive the estimated neutrino mass toward negative values. This is plausible as it is only observed in eBOSS data, which are inherently less constraining, and thus, more susceptible to such prior volume effects. 
Indeed, in the bottom panel of Fig.~\ref{figure:triangular}, we can observe small shifts between the peak of the 1-dimensional distributions and the cosmological parameters used to construct the noiseless synthetic data. Despite these shifts falling within $1\,\sigma$ intervals, one might expect better performance with noiseless data, as recently observed in \cite{Maus:2024sbb, Noriega:2024eyu}, where the distributions match the cosmological parameters, even when using exceedingly large volumes.

\begin{figure}
 	\begin{center}
 	\includegraphics[width=3.0 in]{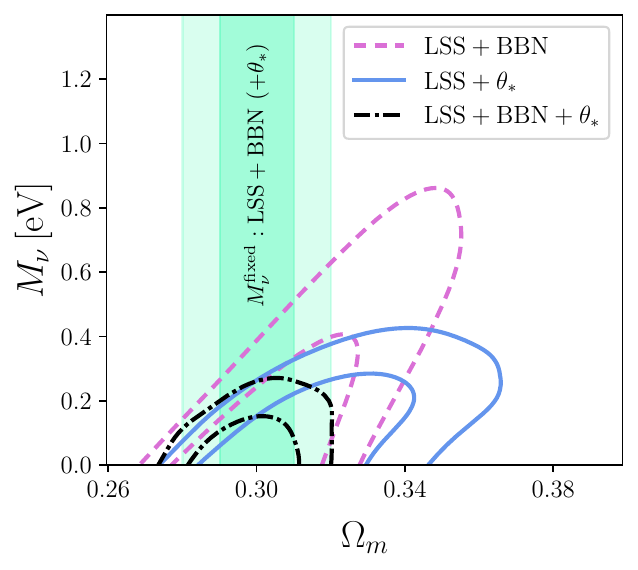}
 	\includegraphics[width=3.0 in]{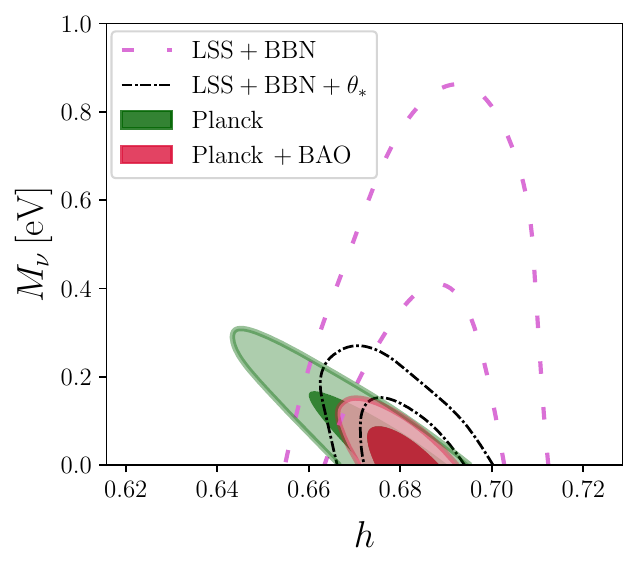}
 	\caption{2-dimensional constraints in the $M_\nu$-$\Omega_m$ space (\textit{top panel}) and $M_\nu$-$h$ space (\textit{bottom panel}). LSS refers to BOSS + eBOSS. Planck (+BAO) refers to the results obtained from CMB data using TT, TE, EE+lowE+lensing (+BAO from BOSS DR12). }
  \label{fig:LSS_Mnu_vs_OmM}
 \end{center}
 \end{figure}

\begin{figure}
 	\begin{center}
 	\includegraphics[width=3.0 in]{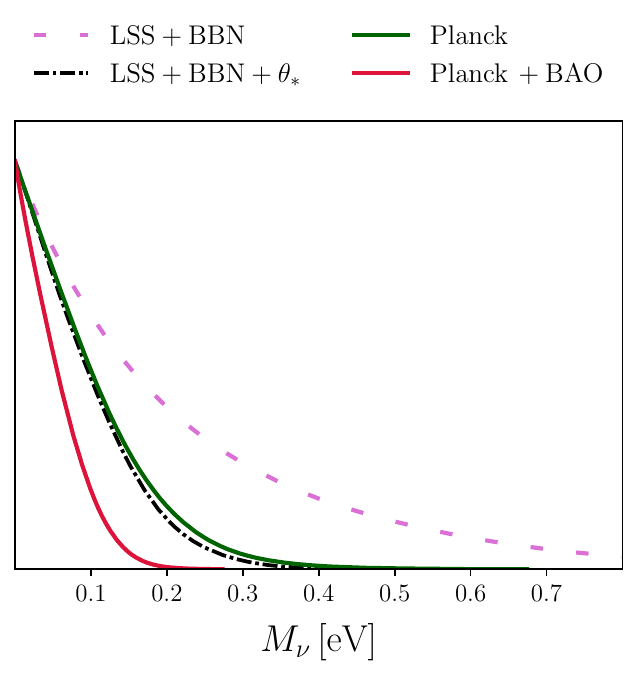}
 	\caption{1-dimensional constraints on $M_\nu$. Labels for LSS and Planck (+BAO) are described in the caption of Fig.~\ref{fig:LSS_Mnu_vs_OmM}.
  }
 \label{fig:Mnu}
 \end{center}
 \end{figure}

Projection effects are typically alleviated with the addition of more data. 
Thus, we incorporate the information from the acoustic scale at the last scattering through the parameter $100 \times \theta_* = 1.04110 \pm 0.00031$ from \textit{Planck 2018}, as a Gaussian prior to our dataset. The outcomes are shown in Fig.~\ref{fig:LSS_Mnu_vs_OmM}, where we can notice that $\theta_*$ contributes to tighter constraints on $\Omega_m$ and $h$, subsequently leading to a more stringent inference of the neutrino mass given by
\begin{align}
&\sumnu< 0.21 \,\text{eV} \quad(95 \,\%, \text{LSS + BBN + $\theta_*$}),     
\end{align}
where LSS refers to the result of jointly fitting the BOSS and eBOSS data. This result achieves slightly better constraints on the neutrino mass compared to those obtained from CMB data by Planck (TT, TE, EE+lowE+lensing) \cite{Planck:2018vyg}. Intermediate results without a $\theta_*$ or BBN prior, as well as those using three degenerate neutrino mass states with physically motivated priors are presented in Table~\ref{table:constraints}. These results show that the upper bound on $M_\nu$ significantly depends on the choice of its prior, whereas the rest of the cosmological parameters are largely unaffected by this prior choice.

In Fig.~\ref{fig:LSS_Mnu_vs_OmM}, several important points are illustrated. In the top panel, it is observed that the additional information provided by BBN and $\theta_*$ have complementary degeneracy directions.  
Therefore, the combined analysis LSS + BBN + $\theta_*$ achieves significantly greater constraining power on $\Omega_m$ and $M_\nu$. Specifically, the constraint on $\Omega_m$ is predominantly influenced by the inclusion of the BBN prior (dashed purple line). 
In contrast, $\theta_*$ does not significantly aid in constraining $\Omega_m$ (blue line) as much as BBN does, but it substantially lowers the upper bound on the neutrino mass. The combination of BBN and $\theta_*$ results in the $M_\nu$--$\Omega_m$ posterior following the profile of $\theta_*$ alone for small $\Omega_m$. It would continue this trend, but BBN helps to tighten the upper bound of $\Omega_m$.

Following with the bottom panel of Fig.~\ref{fig:LSS_Mnu_vs_OmM}, it is observed that the inclusion of $\theta_*$ to LSS+BBN drastically tightens the constraints on $M_\nu$ and recovers the degeneracy direction seen in the Planck results. Despite the high constraining power, the peak at $\sumnu=0$ and potential projection effects remain, as illustrated in Fig.~\ref{fig:Mnu}. In those plots, we also show the posteriors for Planck and Planck + BAO (BOSS DR12) for comparisons.

At this point, one might speculate that the peculiar behavior of the peak at $M_\nu = 0$ could be linked to the tendency of $A_s$ towards lower values compared to Planck, as previously observed in Fig.~\ref{fig:comparison_LCDM} for the $\Lambda$CDM model with fixed neutrino mass. Since, in the case of interest with free neutrino mass, these two parameters are degenerate. However, Fig.~\ref{fig:Mnu_vs_As} shows that the posterior distribution of $M_\nu$ is unaffected by the lower $A_s$ values compared to Planck.

To gain more insight on $M_\nu$ posteriors for eBOSS-like data, we reanalyze the noiseless synthetic data with the covariance matrix rescaled by a factor of $1/25$ (labeled as $\text{cov}_{25}$), corresponding to a volume 25 times larger than that of eBOSS.
In this case, we are able to recover all cosmological parameters with good precision, as shown in Fig.~\ref{fig:cov1_cov25}. However, the neutrino mass profile still appears to have its maximum at negative mass values, though it shows a notably improvement compared to the non-rescaled covariance case (labeled as $\text{cov}_{1}$). Furthermore, the posterior profile of the neutrino mass shows a slight increase in probability density around the expected $M_\nu = 0.06\,\text{eV}$.

\begin{figure}
 	\begin{center}
 	\includegraphics[width=3.3 in]{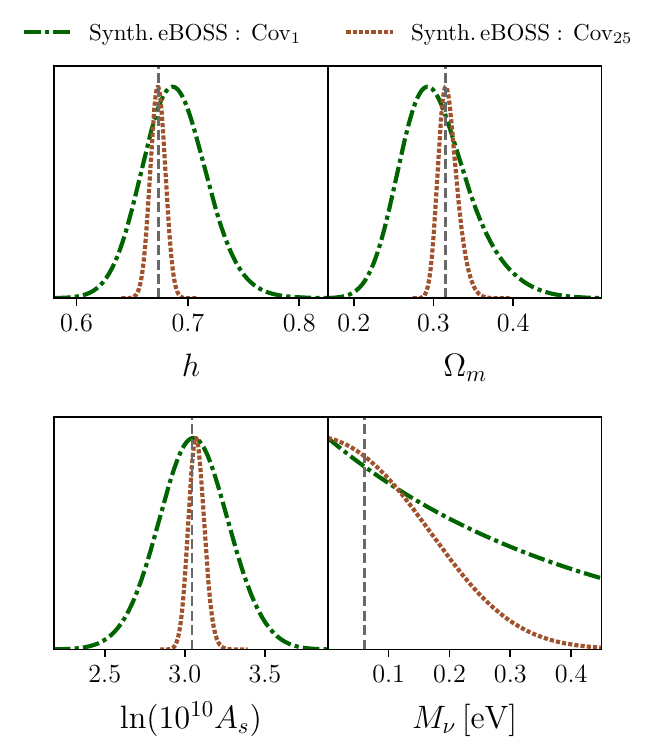}
 	\caption{Constraints for noiseless synthetic eBOSS data using the non-rescaled covariance (cov$_1$) and the $\times 25$ larger volume covariance (cov$_{25}$). Vertical dashed lines represent the best-fit values from \textit{Planck 2018}.
  }
  \label{fig:cov1_cov25}
 \end{center}
 \end{figure}

\begin{figure*}
 	\begin{center}
 	\includegraphics[width=6.4 in]{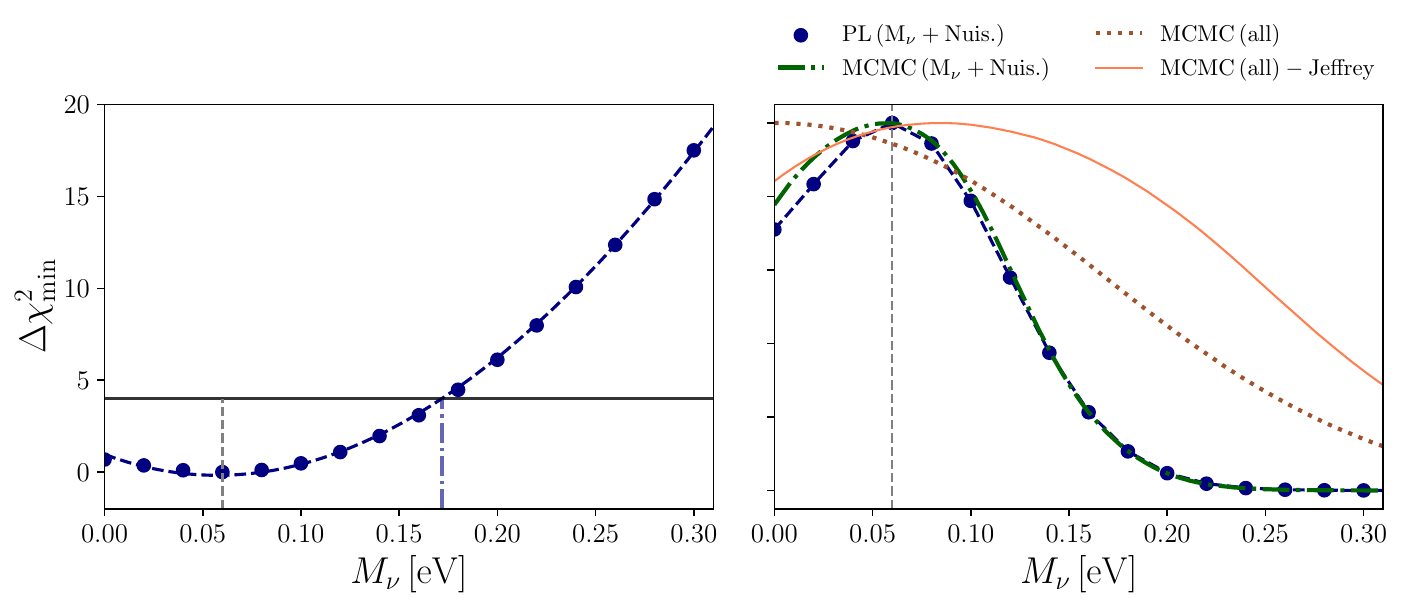}
 	\caption{\textit{Left panel}: $\Delta \chi^2_{\rm min}$ for a set of $M_\nu$ values obtained by profiling the likelihood using the \texttt{iminuit} code on noiseless synthetic eBOSS data with the rescaled covariance matrix $\text{cov}_{25}$. The intersection between the black horizontal line at $\Delta \chi^2_{\rm min} = 3.84$ and the vertical dash-dotted blue line can be used to approximately estimate the upper bound at the 95\% c.l. \cite{Neyman:1937uhy}.
  \textit{Right panel}: Comparison of marginalized posterior distributions obtained from the profile likelihood (PL) and MCMC fits. The dashed blue and dot-dashed green lines correspond to the results of sampling the neutrino mass $\sumnu$ and nuisance parameters (Nuis.) while keeping the other cosmological parameters at their true values. The orange curves (labeled \textit{all}) represent the results obtained from MCMC when fitting $\Lambda$CDM parameters, including the neutrino mass and nuisances. The dotted posterior, already presented in Fig.~\ref{fig:cov1_cov25}, is obtained using the baseline priors introduced in Sect.~\ref{sect:data}, while the solid one corresponds to the results using the Jeffreys prior. In both panels, the vertical dashed gray lines correspond to the maximum \textit{a posteriori} probability (MAP), located by construction at $\sumnu = 0.06\, \text{eV}$.  }
  \label{fig:profile_Mnu}
 \end{center}
 \end{figure*}

A method to identify projection effects consists on comparing results between Bayesian MCMC and the frequentist profile likelihood (PL)  \cite{Cousins:1994yw}, since by construction the latter is unaffected by prior choices and marginalization \cite{Hamann:2007pi, Hadzhiyska:2023wae}. In cosmology this method has been used for Planck CMB data in \cite{Planck:2013nga, Planck:2013pxb}, and more recently it has been used to analyze the galaxy power spectrum full-shape fits \cite{Holm:2023laa, Moretti:2023drg, Herold:2021ksg, Maus:2023rtr, Cruz:2023cxy}. Computing the PL requires the evaluation of the parameters of interest —in our case, $M_\nu$— across a range of values, while maximizing the likelihood $\mathcal{L}$ (or minimizing $\chi^2 = -2 \ln \mathcal{L}$) with respect to all other parameters for each value of $M_\nu$. In this work we obtain the PL using the MINOS algorithm from the \texttt{iminuit} package\footnote{\url{https://scikit-hep.org/iminuit/index.html}} \cite{James:1975dr, iminuit}, by profiling only the neutrino mass parameter. The resulting PL is then expressed as a $\Delta \chi^2_{\rm min}$ distribution as a function of  $M_\nu$.

Figure \ref{fig:profile_Mnu} shows the comparison between Bayesian and frequentist approaches using eBOSS synthetic data and rescaled covariance matrix $\text{cov}_{25}$. Several key points can be identified.  Firstly, the PL analysis confirms that the theoretical model prefers $M_\nu = 0.06\, \text{eV}$, as expected because the data is generated by the same model.

Secondly, the right panel shows similar posterior distributions between the MCMC (dot-dashed green line) and PL (dashed blue line) approaches when profiling the neutrino mass and the nuisance parameters while maintaining the other cosmological parameters at their true values. However, there are still slight differences in the posterior distributions. Notably, the peak of the MCMC analysis is subtly shifted to lower values compared to the expected value. Hence this shift should be the consequence of projection effects arising from the marginalization over nuisance parameters in the Bayesian inference. These effects become more pronounced when extending the sampling space to the rest of the $\Lambda$CDM parameters (labeled \textit{all}). This is evidenced by the orange dotted line, where the peak shifts to $M_\nu = 0$. This suggests that the latter is influenced by projection effects arising not only from the marginalization over nuisance parameters, as observed in \cite{Simon:2022lde, Holm:2023laa} for fixed neutrino mass, but also from the rest of cosmological parameters.
By incorporating the Jeffreys prior, represented by the solid orange line, the profile distribution peaks closer to the expected value $M_\nu = 0.06\, \text{eV}$.
The significant improvement observed when using the Jeffreys prior can be attributed to their ability to partially cancel the contributions from the Laplace term (i.e., $\ln[\det(\mathcal{A}_{ij})]$ in Eq.~(A.9) of \cite{Noriega:2024eyu}), which are directly associated with projection effects \cite{Hadzhiyska:2023wae}. 
\end{section}

\begin{section}{Wiggle vs Broadband suppression }\label{sect:wiggles}
\begin{figure*}
  \centering
  \includegraphics[width=0.45\textwidth, height=0.25\textheight]{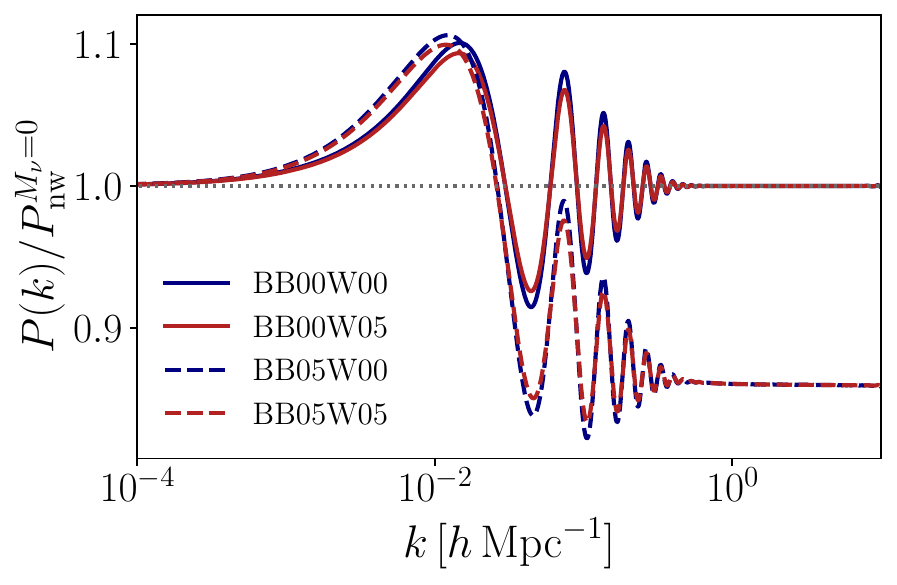}
  \includegraphics[width=0.45\textwidth, height=0.25\textheight]{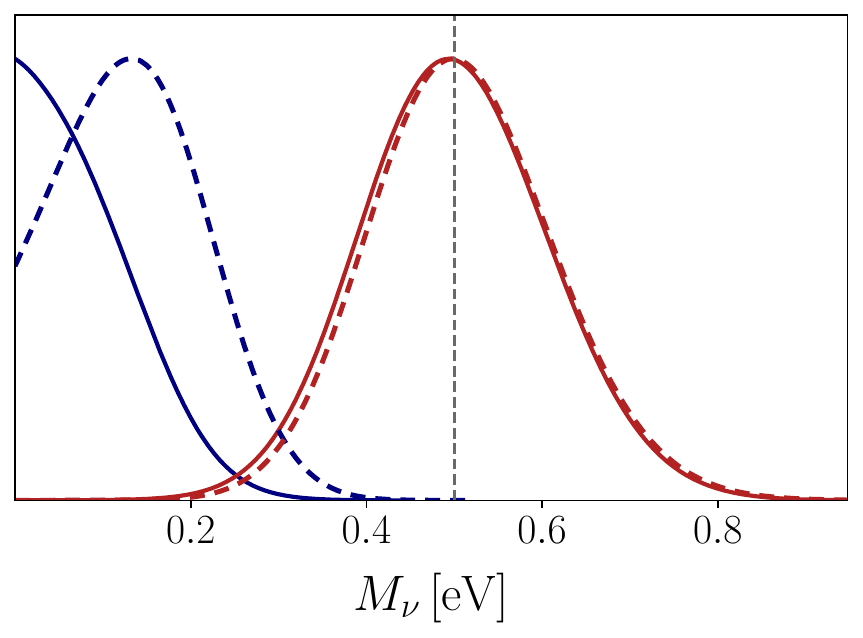}
  \caption{\textit{Frankenstein Experiment}. The \textit{left panel} displays noiseless synthetic data for all possible combinations of broadband (BB) and wiggle (W) features. The \textit{right panel} presents the 1-dimensional marginalized posterior distributions of $M_\nu$ for fits on each synthetic data set, showing that the information on neutrino masses is primarily extracted from the wiggles rather than the broadband suppression.}
  \label{fig:BBvsW}
\end{figure*}

The previous section demonstrates that the current full-shape analysis is not reliable for extracting the neutrino mass from SDSS or even DESI data. Therefore, in this section, we revisit the suppression of the power spectrum to shift the focus away from the effects of neutrinos on background evolution, which can lead to high degeneracies between the neutrino mass and other cosmological parameters.

We first note, that regarding the broadband information of the power spectrum a degeneracy between $\Omega_m$ and $M_\nu$ could be expected, as massive neutrinos suppress the power spectrum, suggesting a compensatory increase in matter density. A similar effect is observed with $A_s$ and $h$, though $A_s$ shows a milder correlation while $h$ is highly sensitive to redshift.\footnote{A similar behavior is observed for $f(R)$ scale-dependent modified gravity (MG) in \cite{Rodriguez-Meza:2023rga}. In that case, however, MG produces an enhancement of the power spectrum instead of a suppression, and $\Omega_m$ and $M_\nu$ become anticorrelated. Furthermore, in viables $f(R)$ scenarios the background evolution is essentially the same as in $\Lambda$CDM.} This leads us to question: where is the information about neutrino mass coming from? Aside from the background evolution, is the neutrino mass inferred from the wiggles rather than broadband information? 
To explore this, we conduct an unorthodox experiment, in the style of Frankenstein's monster \cite{shelley1998}.
We consider two linear power spectra, one with neutrino mass $\sumnu=0.0\,\text{eV}$
and a second with a large neutrino mass, $\sumnu=0.5\,\text{eV}$. We fix the densities $\omega_c$ and $\omega_b$, hence the models have different total amount of matter $\omega_m$ and the effects of massive neutrinos in the wiggles is mainly a modulation in their relative amplitude, with only a very small shift in their frequency. We decompose the two linear power spectra into broadband + wiggles and mix the pieces to get four models. Afterwards, the linear spectra are evolved non-linearly using \texttt{FOLPS}, employing nuisance parameters previously obtained from the best fit to the synthetic data with massless neutrino cosmology, and produce the following four noiseless synthetic data:

\begin{itemize}
    \item B00W00 - Broadband of the model with mass $\sumnu=0.0\,\text{eV}$, wiggles of the model with mass $\sumnu=0.0\,\text{eV}$,
    \item B00W05 - Broadband of the model with mass $\sumnu=0.0\,\text{eV}$, wiggles of the model with mass $\sumnu=0.5\,\text{eV}$, 
    \item B05W00 - Broadband of the model with mass $\sumnu=0.5\,\text{eV}$, wiggles of the model with mass $\sumnu=0.0\,\text{eV}$,
    \item B05W05 - Broadband of the model with mass $\sumnu=0.5\,\text{eV}$, wiggles of the model with mass $\sumnu=0.5\,\text{eV}$, 
\end{itemize}
where B00W00 and B05W05 are used as the original control samples, while the mixed broadband and wiggles correspond to the Frankensteins. For the fits, we utilize the rescaled covariance matrix $\text{cov}_{25}$ and the window function employed for the eBOSS synthetic data.

In our analysis we fit the monopole and quadrupole multipoles using \texttt{FOLPS}. The results are shown in Fig.~\ref{fig:BBvsW}. As can be seen, the models with wiggles for a sum neutrino mass $\sumnu=0.5\,\text{eV}$ produce nearly identical results and both peak at 0.5 eV. This is interesting because the BB00W05 model (solid red line) was obtained from the non-suppressed broadband for massless neutrinos. On the other hand, the two models with wiggles obtained from linear power spectrum neutrinos with $\sumnu=0.0\,\text{eV}$ are consistent with massless neutrinos, even though the BB05W00 model has a highly suppressed broadband.  

From these results we conclude that the relevant information regarding the damping of the power spectrum due to the neutrino mass resides in the wiggles, and the full-shape analysis is blind to the suppression of the broadband. 
\end{section}

\begin{section}{Conclusions}\label{sect:conclusions}
In this work, we revisited the constraints on neutrino mass extracted from the full-shape analysis of BOSS galaxies and eBOSS quasars. Surprisingly, we found that eBOSS places tighter constraints on the neutrino mass compared to the two non-overlapping redshift bins of BOSS, $z_1$ and $z_3$. Moreover, it is responsible for the well-marked peak at $\sumnu = 0$ in the joint analysis of BOSS + eBOSS, a signature also observed in other works using CMB and BAO data \cite{Planck:2018vyg, BOSS:2016wmc, eBOSS:2020yzd, Craig:2024tky, Planck:2013nga}. To further understand the observed behavior, we generated noiseless synthetic data for the flat $\Lambda$CDM model using the \textit{Planck 2018} best-fit values with $\sumnu = 0.06 \,\text{eV}$ as input cosmology. Ideally, one would expect to accurately recover the input cosmological parameters, as the noiseless synthetic data was generated using the same theoretical model and settings. However, we found very similar patterns in the neutrino mass posteriors as those seen in real data, indicating the presence of strong projection effects on the neutrino mass constraints.

We investigated whether incorporating additional constraining data alleviates the impact of projection effects on neutrino constraints. To explore this, we performed a combined analysis of BOSS and eBOSS data, integrating informative priors on baryons from BBN and the acoustic scale at last scattering from \textit{Planck 2018}. Although this analysis achieved high constraining power, putting slightly tighter bounds on neutrino mass $\sumnu < 0.21\, \text{eV}$ (95\% c.l.) compared to the limit of $\sumnu < 0.24\, \text{eV}$ (95\% c.l.) obtained by Planck data alone, the peak at $\sumnu = 0$ and prior volume effects still persisted. %
Furthermore, we compared outcomes from Bayesian MCMC and frequentist profile likelihood approaches. Importantly, the latter is inherently immune to projection effects. Our analysis was conducted in a controlled scenario using noiseless synthetic data that mimicked eBOSS spectra, along with a rescaled covariance matrix reflecting a volume 25 times larger than the employed in eBOSS. Our preliminary test evaluated how nuisance parameters alone affect the constraints on neutrino mass. Subsequently, we analyzed how these constraints are influenced by both nuisance and cosmological parameters. Our findings confirm that neutrino mass constraints derived from Bayesian analysis are influenced by projection effects stemming from nuisance parameters. These results are consistent with previous studies based on the $\Lambda$CDM model with fixed neutrino masses, which found that cosmological parameters are similarly affected \cite{Simon:2022lde, Holm:2023laa}. However, we also discovered that cosmological parameters significantly introduce projection effects on neutrino mass constraints. Furthermore, we observed that utilizing the Jeffreys prior helps to partially mitigate them. 

Our findings highlight the importance of carefully handling systematics in full-shape analyses, especially when aiming to obtain reliable constraints on neutrino mass. This is a crucial consideration for current surveys like DESI and beyond, where poor control of projection effects could lead to incorrect constraints. We claim that these unpleasent outcomes are due to large degeneracies with parameters affected by the background evolution of the Universe. For this reason, we revisited the use of power spectrum suppression to infer the neutrino mass and argue that it should be considered as an alternative approach to extracting information about neutrino mass, as initially proposed in \cite{Hu:1997mj}, rather than relying solely on the cumulative effect of the expansion history of the Universe over the structure formation. We analyze this suppression and, contrary to common belief, found that most clustering information on neutrino mass is derived from the relative amplitude of the BAO wiggles rather than from the broadband suppression. This discovery opens the possibility of developing new analysis methods that rely solely on the wiggles information. Despite providing weaker constraints on neutrino mass, such methods could be more robust than those heavily reliant on background expansion. 
\end{section}

\acknowledgments
This work is partially supported by CONAHCyT grant CBF2023-2024-162 and PAPIIT IG102123.
\appendix

\begin{section}{COMPLEMENTARY RESULTS}

\begin{table*}[htpb!]
\begin{center}
\renewcommand{\arraystretch}{1.5}
\normalsize
\setlength{\tabcolsep}{16pt} 
\begin{tabular}{l c c c }
\hline
\textbf{Model / dataset} & $M_\nu\, [\text{eV}]$ & $h$ & $\Omega_m$
 \vspace{-0.4cm}
 \\
\vspace{-0.2cm}
 & \multicolumn{1}{c}{\rule{2cm}{0.4pt}} & \multicolumn{1}{c}{\rule{2cm}{0.4pt}} & \multicolumn{1}{c}{\rule{2cm}{0.4pt}} 
 \\
 & \multicolumn{1}{c}{95\% c.l.} & \multicolumn{1}{c}{68\% c.l.} & \multicolumn{1}{c}{68\% c.l.} \\
\hline\hline

LSS + BBN   & $ < 0.668$ & $0.686\pm 0.011$ & $0.311^{+0.012}_{-0.020}$\\ 
LSS + $\theta_*$ & $< 0.348$ & $0.640^{+0.020}_{-0.027}$ & $0.321^{+0.015}_{-0.021}$ \\ 
LSS + BBN + $\theta_*$ & $< 0.214$ & $0.6793\pm 0.0073$ & $0.2992\pm 0.0094$\\ 
LSS + BBN + $\theta_*$\, (Deg.; $M_\nu > 0$) & $< 0.245$ & $0.6791\pm 0.0074$ & $0.2993\pm 0.0095$\\ 
LSS + BBN + $\theta_*$\, (Deg.; $M_\nu > 0.058\, \text{eV}$)  & $< 0.274$ & $0.6772\pm 0.0072$ & $0.3006\pm 0.0093$\\ 
LSS + BBN + $\theta_*$\, (Deg.; $M_\nu > 0.10\, \text{eV})$  & $< 0.300$ & $0.6760\pm 0.0072$ & $0.3014\pm 0.0094$\\ 
\hline \hline
\end{tabular}
\end{center}
\caption{Constraints on cosmological parameters obtained by jointly fitting BOSS + eBOSS data in combination with external priors. Results with two-sided error bars refer to the marginalized means and 68\% c.l., while the upper bounds on $M_\nu$ correspond to 95\% limits. The constraints based on a model with three degenerate mass states are labeled as `Deg.' and include physically motivated priors on neutrino mass: $M_\nu > 0$, NH $(M_\nu > 0.058\, \text{eV})$, and IH ($M_\nu > 0.10\, \text{eV}$).} 
\label{table:constraints}
\end{table*}

In this appendix, we present the constraints on cosmological parameters obtained by jointly fitting data from BOSS galaxies and eBOSS quasars, incorporating informative priors from BBN and the acoustic scale $\theta_*$ at last scattering from \textit{Planck 2018} (labeled as LSS+BBN+$\theta_*$), as well as adding only BBN or $\theta_*$ separately. 1-dimensional confidence intervals are shown in Table \ref{table:constraints}. We further put constraints on the neutrino mass for IH and NH, which are obtained by simply putting priors of $\sumnu > 0.10\,\text{eV}$ and  $\sumnu > 0.058\,\text{eV}$, respectively.

Furthermore, Fig.~\ref{fig:Mnu_vs_As} compares the results of directly sampling the bias parameters $b_n = \{b_1, b_2, b_{\text{s}^2}, b_{\text{3nl}}\}$ (dash-dotted black line) against sampling the scaled bias parameters at each step of the MCMC, $b_n \sigma^m_8 = \{b_1\sigma_8, b_2\sigma^2_8, b_{\text{s}^2} \sigma^2_8, b_{\text{3nl}}\sigma^3_8\}$ (solid purple line), following \cite{Maus:2024dzi}, and converting back to the original bias parameters $b_n$ to feed \texttt{FOLPS} and generate the multipoles. This approach helps prevent the downward shift in $A_s$ (and $\sigma_8$). Specifically, we find that the constraints improve from $\ln(10^{10}A_s) = 2.84\pm 0.12$ ($\sigma_8 = 0.710\pm 0.043$) to $\ln(10^{10}A_s) = 2.96\pm 0.12$ ($\sigma_8 = 0.755\pm 0.043$) at 68\% confidence level. However, it does not eliminate the projection effects on $M_\nu$. Indeed, the rest of the cosmological parameters remain unchanged.

\begin{figure}
 	\begin{center}
 	\includegraphics[width=3.4 in]{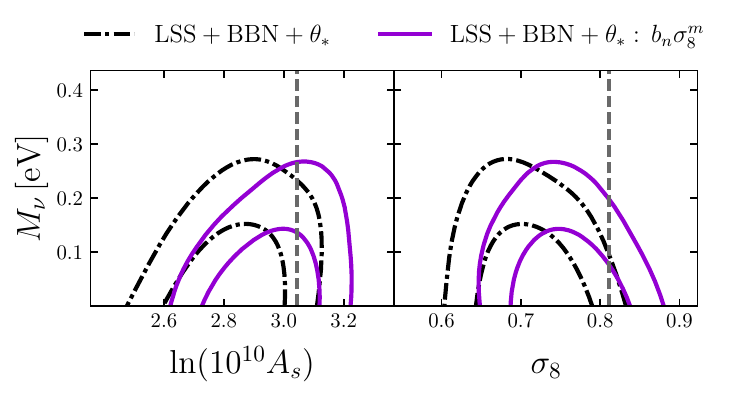}
        \includegraphics[width=3.4 in]{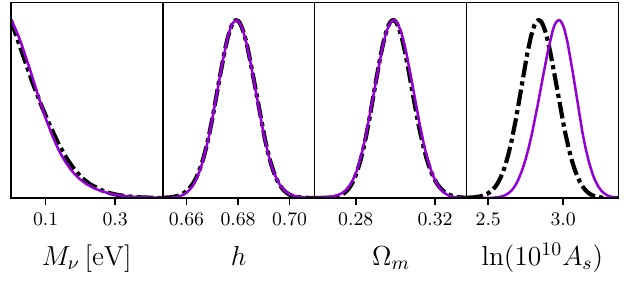}
 	\caption{\textit{Upper panel:} Comparison of 2-dimensional constraints in the $M_\nu$-$\ln(10^{10}A_s)$ and $M_\nu$-$\sigma_8$ spaces for the base setup (dash-dotted black line) and when scaling bias parameters with $\sigma_8$ (solid purple line). The vertical dashed gray lines represent the best-fit values from \textit{Planck 2018}. 
  \textit{Bottom panel:} Comparison of 1-dimensional marginalized posterior distributions of cosmological parameters.
  }
  \label{fig:Mnu_vs_As}
 \end{center}
 \end{figure}
\end{section}

\bibliographystyle{JHEP} 
\bibliography{refs.bib}

\providecommand{\href}[2]{#2}\begingroup\raggedright\begin{thebibliography}{10}

\bibitem{Esteban:2020cvm}
I.~Esteban, M.~C. Gonzalez-Garcia, M.~Maltoni, T.~Schwetz and A.~Zhou, \emph{{The fate of hints: updated global analysis of three-flavor neutrino oscillations}}, \href{http://dx.doi.org/10.1007/JHEP09(2020)178}{\emph{JHEP} {\bf 09} (2020) 178}, [\href{http://arxiv.org/abs/2007.14792}{{\tt 2007.14792}}].

\bibitem{Aker:2024drp}
M.~Aker et~al., \emph{{Direct neutrino-mass measurement based on 259 days of KATRIN data}},  \href{http://arxiv.org/abs/2406.13516}{{\tt 2406.13516}}.

\bibitem{Planck:2018vyg}
{\scshape Planck} collaboration, N.~Aghanim et~al., \emph{{Planck 2018 results. VI. Cosmological parameters}}, \href{http://dx.doi.org/10.1051/0004-6361/201833910}{\emph{Astron. Astrophys.} {\bf 641} (2020) A6}, [\href{http://arxiv.org/abs/1807.06209}{{\tt 1807.06209}}].

\bibitem{DESI:2024mwx}
{\scshape DESI} collaboration, A.~G. Adame et~al., \emph{{DESI 2024 VI: Cosmological Constraints from the Measurements of Baryon Acoustic Oscillations}},  \href{http://arxiv.org/abs/2404.03002}{{\tt 2404.03002}}.

\bibitem{Lesgourgues:2006nd}
J.~Lesgourgues and S.~Pastor, \emph{{Massive neutrinos and cosmology}}, \href{http://dx.doi.org/10.1016/j.physrep.2006.04.001}{\emph{Phys. Rept.} {\bf 429} (2006) 307--379}, [\href{http://arxiv.org/abs/astro-ph/0603494}{{\tt astro-ph/0603494}}].

\bibitem{Wong:2011ip}
Y.~Y.~Y. Wong, \emph{{Neutrino mass in cosmology: status and prospects}}, \href{http://dx.doi.org/10.1146/annurev-nucl-102010-130252}{\emph{Ann. Rev. Nucl. Part. Sci.} {\bf 61} (2011) 69--98}, [\href{http://arxiv.org/abs/1111.1436}{{\tt 1111.1436}}].

\bibitem{BOSS:2016wmc}
{\scshape BOSS} collaboration, S.~Alam et~al., \emph{{The clustering of galaxies in the completed SDSS-III Baryon Oscillation Spectroscopic Survey: cosmological analysis of the DR12 galaxy sample}}, \href{http://dx.doi.org/10.1093/mnras/stx721}{\emph{Mon. Not. Roy. Astron. Soc.} {\bf 470} (2017) 2617--2652}, [\href{http://arxiv.org/abs/1607.03155}{{\tt 1607.03155}}].

\bibitem{eBOSS:2020yzd}
{\scshape eBOSS} collaboration, S.~Alam et~al., \emph{{Completed SDSS-IV extended Baryon Oscillation Spectroscopic Survey: Cosmological implications from two decades of spectroscopic surveys at the Apache Point Observatory}}, \href{http://dx.doi.org/10.1103/PhysRevD.103.083533}{\emph{Phys. Rev. D} {\bf 103} (2021) 083533}, [\href{http://arxiv.org/abs/2007.08991}{{\tt 2007.08991}}].

\bibitem{Craig:2024tky}
N.~Craig, D.~Green, J.~Meyers and S.~Rajendran, \emph{{No $\nu$s is Good News}},  \href{http://arxiv.org/abs/2405.00836}{{\tt 2405.00836}}.

\bibitem{Aviles:2021que}
A.~Aviles, A.~Banerjee, G.~Niz and Z.~Slepian, \emph{{Clustering in massive neutrino cosmologies via Eulerian Perturbation Theory}}, \href{http://dx.doi.org/10.1088/1475-7516/2021/11/028}{\emph{JCAP} {\bf 11} (2021) 028}, [\href{http://arxiv.org/abs/2106.13771}{{\tt 2106.13771}}].

\bibitem{Moretti:2023drg}
C.~Moretti, M.~Tsedrik, P.~Carrilho and A.~Pourtsidou, \emph{{Modified gravity and massive neutrinos: constraints from the full shape analysis of BOSS galaxies and forecasts for Stage IV surveys}}, \href{http://dx.doi.org/10.1088/1475-7516/2023/12/025}{\emph{JCAP} {\bf 12} (2023) 025}, [\href{http://arxiv.org/abs/2306.09275}{{\tt 2306.09275}}].

\bibitem{2013AJ....145...10DBOSS:2016wmc}
K.~S. {Dawson}, D.~J. {Schlegel}, C.~P. {Ahn}, S.~F. {Anderson}, {\'E}.~{Aubourg}, S.~{Bailey} et~al., \emph{{The Baryon Oscillation Spectroscopic Survey of SDSS-III}}, \href{http://dx.doi.org/10.1088/0004-6256/145/1/10}{\emph{\aj} {\bf 145} (Jan., 2013) 10}, [\href{http://arxiv.org/abs/1208.0022}{{\tt 1208.0022}}].

\bibitem{Dawson:2015wdb}
K.~S. Dawson et~al., \emph{{The SDSS-IV extended Baryon Oscillation Spectroscopic Survey: Overview and Early Data}}, \href{http://dx.doi.org/10.3847/0004-6256/151/2/44}{\emph{Astron. J.} {\bf 151} (2016) 44}, [\href{http://arxiv.org/abs/1508.04473}{{\tt 1508.04473}}].

\bibitem{Bernardeau:2001qr}
F.~Bernardeau, S.~Colombi, E.~Gaztanaga and R.~Scoccimarro, \emph{{Large scale structure of the universe and cosmological perturbation theory}}, \href{http://dx.doi.org/10.1016/S0370-1573(02)00135-7}{\emph{Phys. Rept.} {\bf 367} (2002) 1--248}, [\href{http://arxiv.org/abs/astro-ph/0112551}{{\tt astro-ph/0112551}}].

\bibitem{McDonald:2009dh}
P.~McDonald and A.~Roy, \emph{{Clustering of dark matter tracers: generalizing bias for the coming era of precision LSS}}, \href{http://dx.doi.org/10.1088/1475-7516/2009/08/020}{\emph{JCAP} {\bf 0908} (2009) 020}, [\href{http://arxiv.org/abs/0902.0991}{{\tt 0902.0991}}].

\bibitem{Baumann:2010tm}
D.~Baumann, A.~Nicolis, L.~Senatore and M.~Zaldarriaga, \emph{{Cosmological Non-Linearities as an Effective Fluid}}, \href{http://dx.doi.org/10.1088/1475-7516/2012/07/051}{\emph{JCAP} {\bf 07} (2012) 051}, [\href{http://arxiv.org/abs/1004.2488}{{\tt 1004.2488}}].

\bibitem{Vlah:2015sea}
Z.~Vlah, M.~White and A.~Aviles, \emph{{A Lagrangian effective field theory}}, \href{http://dx.doi.org/10.1088/1475-7516/2015/09/014}{\emph{JCAP} {\bf 09} (2015) 014}, [\href{http://arxiv.org/abs/1506.05264}{{\tt 1506.05264}}].

\bibitem{Ivanov:2019pdj}
M.~M. Ivanov, M.~Simonovi\'c and M.~Zaldarriaga, \emph{{Cosmological Parameters from the BOSS Galaxy Power Spectrum}}, \href{http://dx.doi.org/10.1088/1475-7516/2020/05/042}{\emph{JCAP} {\bf 05} (2020) 042}, [\href{http://arxiv.org/abs/1909.05277}{{\tt 1909.05277}}].

\bibitem{DAmico:2019fhj}
G.~D'Amico, J.~Gleyzes, N.~Kokron, K.~Markovic, L.~Senatore, P.~Zhang et~al., \emph{{The Cosmological Analysis of the SDSS/BOSS data from the Effective Field Theory of Large-Scale Structure}}, \href{http://dx.doi.org/10.1088/1475-7516/2020/05/005}{\emph{JCAP} {\bf 05} (2020) 005}, [\href{http://arxiv.org/abs/1909.05271}{{\tt 1909.05271}}].

\bibitem{Wadekar:2020hax}
D.~Wadekar, M.~M. Ivanov and R.~Scoccimarro, \emph{{Cosmological constraints from BOSS with analytic covariance matrices}}, \href{http://dx.doi.org/10.1103/PhysRevD.102.123521}{\emph{Phys. Rev. D} {\bf 102} (2020) 123521}, [\href{http://arxiv.org/abs/2009.00622}{{\tt 2009.00622}}].

\bibitem{Chudaykin:2020aoj}
A.~Chudaykin, M.~M. Ivanov, O.~H. Philcox and M.~Simonovi\'c, \emph{{Nonlinear perturbation theory extension of the Boltzmann code CLASS}}, \href{http://dx.doi.org/10.1103/PhysRevD.102.063533}{\emph{Phys. Rev. D} {\bf 102} (2020) 063533}, [\href{http://arxiv.org/abs/2004.10607}{{\tt 2004.10607}}].

\bibitem{Philcox:2021kcw}
O.~H.~E. Philcox and M.~M. Ivanov, \emph{{BOSS DR12 full-shape cosmology: \ensuremath{\Lambda}CDM constraints from the large-scale galaxy power spectrum and bispectrum monopole}}, \href{http://dx.doi.org/10.1103/PhysRevD.105.043517}{\emph{Phys. Rev. D} {\bf 105} (2022) 043517}, [\href{http://arxiv.org/abs/2112.04515}{{\tt 2112.04515}}].

\bibitem{Philcox:2022frc}
O.~H.~E. Philcox, M.~M. Ivanov, G.~Cabass, M.~Simonovi\'c, M.~Zaldarriaga and T.~Nishimichi, \emph{{Cosmology with the Redshift-Space Galaxy Bispectrum Monopole at One-Loop Order}},  \href{http://arxiv.org/abs/2206.02800}{{\tt 2206.02800}}.

\bibitem{Tanseri:2022zfe}
I.~Tanseri, S.~Hagstotz, S.~Vagnozzi, E.~Giusarma and K.~Freese, \emph{{Updated neutrino mass constraints from galaxy clustering and CMB lensing-galaxy cross-correlation measurements}},  \href{http://arxiv.org/abs/2207.01913}{{\tt 2207.01913}}.

\bibitem{Nishimichi:2020tvu}
T.~Nishimichi, G.~D'Amico, M.~M. Ivanov, L.~Senatore, M.~Simonovi\'c, M.~Takada et~al., \emph{{Blinded challenge for precision cosmology with large-scale structure: results from effective field theory for the redshift-space galaxy power spectrum}}, \href{http://dx.doi.org/10.1103/PhysRevD.102.123541}{\emph{Phys. Rev. D} {\bf 102} (2020) 123541}, [\href{http://arxiv.org/abs/2003.08277}{{\tt 2003.08277}}].

\bibitem{Chen:2020zjt}
S.-F. Chen, Z.~Vlah, E.~Castorina and M.~White, \emph{{Redshift-Space Distortions in Lagrangian Perturbation Theory}}, \href{http://dx.doi.org/10.1088/1475-7516/2021/03/100}{\emph{JCAP} {\bf 03} (2021) 100}, [\href{http://arxiv.org/abs/2012.04636}{{\tt 2012.04636}}].

\bibitem{Tsedrik:2022cri}
M.~Tsedrik, C.~Moretti, P.~Carrilho, F.~Rizzo and A.~Pourtsidou, \emph{{Interacting dark energy from the joint analysis of the power spectrum and bispectrum multipoles with the EFTofLSS}},  \href{http://arxiv.org/abs/2207.13011}{{\tt 2207.13011}}.

\bibitem{Carrilho:2022mon}
P.~Carrilho, C.~Moretti and A.~Pourtsidou, \emph{{Cosmology with the EFTofLSS and BOSS: dark energy constraints and a note on priors}},  \href{http://arxiv.org/abs/2207.14784}{{\tt 2207.14784}}.

\bibitem{Nunes:2022bhn}
R.~C. Nunes, S.~Vagnozzi, S.~Kumar, E.~Di~Valentino and O.~Mena, \emph{{New tests of dark sector interactions from the full-shape galaxy power spectrum}}, \href{http://dx.doi.org/10.1103/PhysRevD.105.123506}{\emph{Phys. Rev. D} {\bf 105} (2022) 123506}, [\href{http://arxiv.org/abs/2203.08093}{{\tt 2203.08093}}].

\bibitem{Ramirez:2023ads}
S.~Ramirez, M.~Icaza-Lizaola, S.~Fromenteau, M.~Vargas-Maga\~na and A.~Aviles, \emph{{Full Shape Cosmology Analysis from BOSS in configuration space using Neural Network Acceleration}},  \href{http://arxiv.org/abs/2310.17834}{{\tt 2310.17834}}.

\bibitem{Chen:2020fxs}
S.-F. Chen, Z.~Vlah and M.~White, \emph{{Consistent Modeling of Velocity Statistics and Redshift-Space Distortions in One-Loop Perturbation Theory}}, \href{http://dx.doi.org/10.1088/1475-7516/2020/07/062}{\emph{JCAP} {\bf 07} (2020) 062}, [\href{http://arxiv.org/abs/2005.00523}{{\tt 2005.00523}}].

\bibitem{DAmico:2020kxu}
G.~D'Amico, L.~Senatore and P.~Zhang, \emph{{Limits on $w$CDM from the EFTofLSS with the PyBird code}}, \href{http://dx.doi.org/10.1088/1475-7516/2021/01/006}{\emph{JCAP} {\bf 01} (2021) 006}, [\href{http://arxiv.org/abs/2003.07956}{{\tt 2003.07956}}].

\bibitem{Linde:2024uzr}
D.~Linde, A.~Moradinezhad~Dizgah, C.~Radermacher, S.~Casas and J.~Lesgourgues, \emph{{CLASS-OneLoop: Accurate and Unbiased Inference from Spectroscopic Galaxy Surveys}},  \href{http://arxiv.org/abs/2402.09778}{{\tt 2402.09778}}.

\bibitem{Noriega:2022nhf}
H.~E. Noriega, A.~Aviles, S.~Fromenteau and M.~Vargas-Maga\~na, \emph{{Fast computation of non-linear power spectrum in cosmologies with massive neutrinos}}, \href{http://dx.doi.org/10.1088/1475-7516/2022/11/038}{\emph{JCAP} {\bf 11} (2022) 038}, [\href{http://arxiv.org/abs/2208.02791}{{\tt 2208.02791}}].

\bibitem{Aviles:2020cax}
A.~Aviles and A.~Banerjee, \emph{{A Lagrangian Perturbation Theory in the presence of massive neutrinos}}, \href{http://dx.doi.org/10.1088/1475-7516/2020/10/034}{\emph{JCAP} {\bf 10} (2020) 034}, [\href{http://arxiv.org/abs/2007.06508}{{\tt 2007.06508}}].

\bibitem{Rodriguez-Meza:2023rga}
M.~A. Rodriguez-Meza, A.~Aviles, H.~E. Noriega, C.-Z. Ruan, B.~Li, M.~Vargas-Maga\~na et~al., \emph{{fkPT: constraining scale-dependent modified gravity with the full-shape galaxy power spectrum}}, \href{http://dx.doi.org/10.1088/1475-7516/2024/03/049}{\emph{JCAP} {\bf 03} (2024) 049}, [\href{http://arxiv.org/abs/2312.10510}{{\tt 2312.10510}}].

\bibitem{Noriega:2024eyu}
H.~E. Noriega et~al., \emph{{Comparing Compressed and Full-modeling Analyses with FOLPS: Implications for DESI 2024 and beyond}},  \href{http://arxiv.org/abs/2404.07269}{{\tt 2404.07269}}.

\bibitem{Kaiser:1984sw}
N.~Kaiser, \emph{{On the Spatial correlations of Abell clusters}}, \href{http://dx.doi.org/10.1086/184341}{\emph{Astrophys. J. Lett.} {\bf 284} (1984) L9--L12}.

\bibitem{Scoccimarro:2004tg}
R.~Scoccimarro, \emph{{Redshift-space distortions, pairwise velocities and nonlinearities}}, \href{http://dx.doi.org/10.1103/PhysRevD.70.083007}{\emph{Phys. Rev.} {\bf D70} (2004) 083007}, [\href{http://arxiv.org/abs/astro-ph/0407214}{{\tt astro-ph/0407214}}].

\bibitem{Taruya:2010mx}
A.~Taruya, T.~Nishimichi and S.~Saito, \emph{{Baryon Acoustic Oscillations in 2D: Modeling Redshift-space Power Spectrum from Perturbation Theory}}, \href{http://dx.doi.org/10.1103/PhysRevD.82.063522}{\emph{Phys. Rev.} {\bf D82} (2010) 063522}, [\href{http://arxiv.org/abs/1006.0699}{{\tt 1006.0699}}].

\bibitem{Aviles:2020wme}
A.~Aviles, G.~Valogiannis, M.~A. Rodriguez-Meza, J.~L. Cervantes-Cota, B.~Li and R.~Bean, \emph{{Redshift space power spectrum beyond Einstein-de Sitter kernels}}, \href{http://dx.doi.org/10.1088/1475-7516/2021/04/039}{\emph{JCAP} {\bf 04} (2021) 039}, [\href{http://arxiv.org/abs/2012.05077}{{\tt 2012.05077}}].

\bibitem{Perko:2016puo}
A.~Perko, L.~Senatore, E.~Jennings and R.~H. Wechsler, \emph{{Biased Tracers in Redshift Space in the EFT of Large-Scale Structure}},  \href{http://arxiv.org/abs/1610.09321}{{\tt 1610.09321}}.

\bibitem{Vlah:2018ygt}
Z.~Vlah and M.~White, \emph{{Exploring redshift-space distortions in large-scale structure}}, \href{http://dx.doi.org/10.1088/1475-7516/2019/03/007}{\emph{JCAP} {\bf 1903} (2019) 007}, [\href{http://arxiv.org/abs/1812.02775}{{\tt 1812.02775}}].

\bibitem{Assassi:2014fva}
V.~Assassi, D.~Baumann, D.~Green and M.~Zaldarriaga, \emph{{Renormalized Halo Bias}}, \href{http://dx.doi.org/10.1088/1475-7516/2014/08/056}{\emph{JCAP} {\bf 08} (2014) 056}, [\href{http://arxiv.org/abs/1402.5916}{{\tt 1402.5916}}].

\bibitem{Chen:2021wdi}
S.-F. Chen, Z.~Vlah and M.~White, \emph{{A new analysis of galaxy 2-point functions in the BOSS survey, including full-shape information and post-reconstruction BAO}}, \href{http://dx.doi.org/10.1088/1475-7516/2022/02/008}{\emph{JCAP} {\bf 02} (2022) 008}, [\href{http://arxiv.org/abs/2110.05530}{{\tt 2110.05530}}].

\bibitem{Simon:2022lde}
T.~Simon, P.~Zhang, V.~Poulin and T.~L. Smith, \emph{{Consistency of effective field theory analyses of the BOSS power spectrum}}, \href{http://dx.doi.org/10.1103/PhysRevD.107.123530}{\emph{Phys. Rev. D} {\bf 107} (2023) 123530}, [\href{http://arxiv.org/abs/2208.05929}{{\tt 2208.05929}}].

\bibitem{Maus:2024sbb}
M.~Maus et~al., \emph{{A comparison of effective field theory models of redshift space galaxy power spectra for DESI 2024 and future surveys}},  \href{http://arxiv.org/abs/2404.07272}{{\tt 2404.07272}}.

\bibitem{Troster:2019ean}
T.~Tr\"oster et~al., \emph{{Cosmology from large-scale structure: Constraining $\Lambda$CDM with BOSS}}, \href{http://dx.doi.org/10.1051/0004-6361/201936772}{\emph{Astron. Astrophys.} {\bf 633} (2020) L10}, [\href{http://arxiv.org/abs/1909.11006}{{\tt 1909.11006}}].

\bibitem{DiValentino:2020vvd}
E.~Di~Valentino et~al., \emph{{Cosmology Intertwined III: $f \sigma_8$ and $S_8$}}, \href{http://dx.doi.org/10.1016/j.astropartphys.2021.102604}{\emph{Astropart. Phys.} {\bf 131} (2021) 102604}, [\href{http://arxiv.org/abs/2008.11285}{{\tt 2008.11285}}].

\bibitem{Abdalla:2022yfr}
E.~Abdalla et~al., \emph{{Cosmology intertwined: A review of the particle physics, astrophysics, and cosmology associated with the cosmological tensions and anomalies}}, \href{http://dx.doi.org/10.1016/j.jheap.2022.04.002}{\emph{JHEAp} {\bf 34} (2022) 49--211}, [\href{http://arxiv.org/abs/2203.06142}{{\tt 2203.06142}}].

\bibitem{Chen:2024vuf}
S.-F. Chen, M.~M. Ivanov, O.~H.~E. Philcox and L.~Wenzl, \emph{{Suppression without Thawing: Constraining Structure Formation and Dark Energy with Galaxy Clustering}},  \href{http://arxiv.org/abs/2406.13388}{{\tt 2406.13388}}.

\bibitem{Maus:2024dzi}
M.~Maus et~al., \emph{{An analysis of parameter compression and full-modeling techniques with Velocileptors for DESI 2024 and beyond}},  \href{http://arxiv.org/abs/2404.07312}{{\tt 2404.07312}}.

\bibitem{Beutler:2021eqq}
F.~Beutler and P.~McDonald, \emph{{Unified galaxy power spectrum measurements from 6dFGS, BOSS, and eBOSS}}, \href{http://dx.doi.org/10.1088/1475-7516/2021/11/031}{\emph{JCAP} {\bf 11} (2021) 031}, [\href{http://arxiv.org/abs/2106.06324}{{\tt 2106.06324}}].

\bibitem{Kitaura:2015uqa}
F.-S. Kitaura et~al., \emph{{The clustering of galaxies in the SDSS-III Baryon Oscillation Spectroscopic Survey: mock galaxy catalogues for the BOSS Final Data Release}}, \href{http://dx.doi.org/10.1093/mnras/stv2826}{\emph{Mon. Not. Roy. Astron. Soc.} {\bf 456} (2016) 4156--4173}, [\href{http://arxiv.org/abs/1509.06400}{{\tt 1509.06400}}].

\bibitem{eBOSS:2020wwo}
{\scshape eBOSS} collaboration, C.~Zhao et~al., \emph{{The completed SDSS-IV extended Baryon Oscillation Spectroscopic Survey: 1000 multi-tracer mock catalogues with redshift evolution and systematics for galaxies and quasars of the final data release}}, \href{http://dx.doi.org/10.1093/mnras/stab510}{\emph{Mon. Not. Roy. Astron. Soc.} {\bf 503} (2021) 1149--1173}, [\href{http://arxiv.org/abs/2007.08997}{{\tt 2007.08997}}].

\bibitem{Aver:2015iza}
E.~Aver, K.~A. Olive and E.~D. Skillman, \emph{{The effects of He I \ensuremath{\lambda}10830 on helium abundance determinations}}, \href{http://dx.doi.org/10.1088/1475-7516/2015/07/011}{\emph{JCAP} {\bf 07} (2015) 011}, [\href{http://arxiv.org/abs/1503.08146}{{\tt 1503.08146}}].

\bibitem{Cooke:2017cwo}
R.~J. Cooke, M.~Pettini and C.~C. Steidel, \emph{{One Percent Determination of the Primordial Deuterium Abundance}}, \href{http://dx.doi.org/10.3847/1538-4357/aaab53}{\emph{Astrophys. J.} {\bf 855} (2018) 102}, [\href{http://arxiv.org/abs/1710.11129}{{\tt 1710.11129}}].

\bibitem{Chan:2012jj}
K.~C. Chan, R.~Scoccimarro and R.~K. Sheth, \emph{{Gravity and Large-Scale Non-local Bias}}, \href{http://dx.doi.org/10.1103/PhysRevD.85.083509}{\emph{Phys. Rev. D} {\bf 85} (2012) 083509}, [\href{http://arxiv.org/abs/1201.3614}{{\tt 1201.3614}}].

\bibitem{Baldauf:2012hs}
T.~Baldauf, U.~Seljak, V.~Desjacques and P.~McDonald, \emph{{Evidence for Quadratic Tidal Tensor Bias from the Halo Bispectrum}}, \href{http://dx.doi.org/10.1103/PhysRevD.86.083540}{\emph{Phys. Rev. D} {\bf 86} (2012) 083540}, [\href{http://arxiv.org/abs/1201.4827}{{\tt 1201.4827}}].

\bibitem{Saito:2014qha}
S.~Saito, T.~Baldauf, Z.~Vlah, U.~Seljak, T.~Okumura and P.~McDonald, \emph{{Understanding higher-order nonlocal halo bias at large scales by combining the power spectrum with the bispectrum}}, \href{http://dx.doi.org/10.1103/PhysRevD.90.123522}{\emph{Phys. Rev.} {\bf D90} (2014) 123522}, [\href{http://arxiv.org/abs/1405.1447}{{\tt 1405.1447}}].

\bibitem{Blas:2011rf}
D.~Blas, J.~Lesgourgues and T.~Tram, \emph{{The Cosmic Linear Anisotropy Solving System (CLASS) II: Approximation schemes}}, \href{http://dx.doi.org/10.1088/1475-7516/2011/07/034}{\emph{JCAP} {\bf 07} (2011) 034}, [\href{http://arxiv.org/abs/1104.2933}{{\tt 1104.2933}}].

\bibitem{ForemanMackey:2012ig}
D.~Foreman-Mackey, D.~W. Hogg, D.~Lang and J.~Goodman, \emph{{emcee: The MCMC Hammer}}, \href{http://dx.doi.org/10.1086/670067}{\emph{Publ. Astron. Soc. Pac.} {\bf 125} (2013) 306--312}, [\href{http://arxiv.org/abs/1202.3665}{{\tt 1202.3665}}].

\bibitem{Lewis:2019xzd}
A.~Lewis, \emph{{GetDist: a Python package for analysing Monte Carlo samples}},  \href{http://arxiv.org/abs/1910.13970}{{\tt 1910.13970}}.

\bibitem{Simon:2022csv}
T.~Simon, P.~Zhang and V.~Poulin, \emph{{Cosmological inference from the EFTofLSS: the eBOSS QSO full-shape analysis}}, \href{http://dx.doi.org/10.1088/1475-7516/2023/07/041}{\emph{JCAP} {\bf 07} (2023) 041}, [\href{http://arxiv.org/abs/2210.14931}{{\tt 2210.14931}}].

\bibitem{Neyman:1937uhy}
J.~Neyman, \emph{{Outline of a Theory of Statistical Estimation Based on the Classical Theory of Probability}}, \href{http://dx.doi.org/10.1098/rsta.1937.0005}{\emph{Phil. Trans. Roy. Soc. Lond. A} {\bf 236} (1937) 333--380}.

\bibitem{Cousins:1994yw}
R.~D. Cousins, \emph{{Why isn't every physicist a Bayesian?}}, \href{http://dx.doi.org/10.1119/1.17901}{\emph{Am. J. Phys.} {\bf 63} (1995) 398}.

\bibitem{Hamann:2007pi}
J.~Hamann, S.~Hannestad, G.~G. Raffelt and Y.~Y.~Y. Wong, \emph{{Observational bounds on the cosmic radiation density}}, \href{http://dx.doi.org/10.1088/1475-7516/2007/08/021}{\emph{JCAP} {\bf 08} (2007) 021}, [\href{http://arxiv.org/abs/0705.0440}{{\tt 0705.0440}}].

\bibitem{Hadzhiyska:2023wae}
B.~Hadzhiyska, K.~Wolz, S.~Azzoni, D.~Alonso, C.~Garc\'\i{}a-Garc\'\i{}a, J.~Ruiz-Zapatero et~al., \emph{{Cosmology with 6 parameters in the Stage-IV era: efficient marginalisation over nuisance parameters}},  \href{http://arxiv.org/abs/2301.11895}{{\tt 2301.11895}}.

\bibitem{Planck:2013nga}
{\scshape Planck} collaboration, P.~A.~R. Ade et~al., \emph{{Planck intermediate results. XVI. Profile likelihoods for cosmological parameters}}, \href{http://dx.doi.org/10.1051/0004-6361/201323003}{\emph{Astron. Astrophys.} {\bf 566} (2014) A54}, [\href{http://arxiv.org/abs/1311.1657}{{\tt 1311.1657}}].

\bibitem{Planck:2013pxb}
{\scshape Planck} collaboration, P.~A.~R. Ade et~al., \emph{{Planck 2013 results. XVI. Cosmological parameters}}, \href{http://dx.doi.org/10.1051/0004-6361/201321591}{\emph{Astron. Astrophys.} {\bf 571} (2014) A16}, [\href{http://arxiv.org/abs/1303.5076}{{\tt 1303.5076}}].

\bibitem{Holm:2023laa}
E.~B. Holm, L.~Herold, T.~Simon, E.~G.~M. Ferreira, S.~Hannestad, V.~Poulin et~al., \emph{{Bayesian and frequentist investigation of prior effects in EFT of LSS analyses of full-shape BOSS and eBOSS data}}, \href{http://dx.doi.org/10.1103/PhysRevD.108.123514}{\emph{Phys. Rev. D} {\bf 108} (2023) 123514}, [\href{http://arxiv.org/abs/2309.04468}{{\tt 2309.04468}}].

\bibitem{Herold:2021ksg}
L.~Herold, E.~G.~M. Ferreira and E.~Komatsu, \emph{{New Constraint on Early Dark Energy from Planck and BOSS Data Using the Profile Likelihood}}, \href{http://dx.doi.org/10.3847/2041-8213/ac63a3}{\emph{Astrophys. J. Lett.} {\bf 929} (2022) L16}, [\href{http://arxiv.org/abs/2112.12140}{{\tt 2112.12140}}].

\bibitem{Maus:2023rtr}
M.~Maus, S.-F. Chen and M.~White, \emph{{A comparison of template vs. direct model fitting for redshift-space distortions in BOSS}}, \href{http://dx.doi.org/10.1088/1475-7516/2023/06/005}{\emph{JCAP} {\bf 06} (2023) 005}, [\href{http://arxiv.org/abs/2302.07430}{{\tt 2302.07430}}].

\bibitem{Cruz:2023cxy}
J.~S. Cruz, S.~Hannestad, E.~B. Holm, F.~Niedermann, M.~S. Sloth and T.~Tram, \emph{{Profiling cold new early dark energy}}, \href{http://dx.doi.org/10.1103/PhysRevD.108.023518}{\emph{Phys. Rev. D} {\bf 108} (2023) 023518}, [\href{http://arxiv.org/abs/2302.07934}{{\tt 2302.07934}}].

\bibitem{James:1975dr}
F.~James and M.~Roos, \emph{{Minuit: A System for Function Minimization and Analysis of the Parameter Errors and Correlations}}, \href{http://dx.doi.org/10.1016/0010-4655(75)90039-9}{\emph{Comput. Phys. Commun.} {\bf 10} (1975) 343--367}.

\bibitem{iminuit}
H.~Dembinski and P.~O. et~al., \emph{scikit-hep/iminuit}, .

\bibitem{shelley1998}
M.~W. Shelley, \emph{Frankenstein, or, The Modern Prometheus: the 1818 Text}.
\newblock Oxford University Press, Oxford; New York, 1998.

\bibitem{Hu:1997mj}
W.~Hu, D.~J. Eisenstein and M.~Tegmark, \emph{{Weighing neutrinos with galaxy surveys}}, \href{http://dx.doi.org/10.1103/PhysRevLett.80.5255}{\emph{Phys. Rev. Lett.} {\bf 80} (1998) 5255--5258}, [\href{http://arxiv.org/abs/astro-ph/9712057}{{\tt astro-ph/9712057}}].

\end{thebibliography}\endgroup

\end{document}